%% file: main.tex
\documentclass[conference]{IEEEtran}
\IEEEoverridecommandlockouts

\input{preamble}

\begin{document}
\bstctlcite{IEEEexample:BSTcontrol} 

\title{Design and Channel Modeling of Electromagnetically Reconfigurable Antennas\\ 
\author{
Ruiqi Wang\IEEEauthorrefmark{1}, 
Pinjun Zheng\IEEEauthorrefmark{2},
Tareq Y. Al-Naffouri\IEEEauthorrefmark{1},
Atif Shamim\IEEEauthorrefmark{1}\\
\textit{\IEEEauthorrefmark{1}King Abdullah University of Science and Technology, KSA}\quad
\textit{\IEEEauthorrefmark{2}University of British Columbia, Canada}\\
(Email: ruiqi.wang.1@kaust.edu.sa \vspace{-0.5em})
} 
}

\maketitle

\begin{abstract}
 
In this work, a novel design of electromagnetically reconfigurable antennas (ERAs) based on a fluid antenna system (FAS) is proposed, and the corresponding wireless channel model is established. Different from conventional antenna arrays with static elements, the electromagnetic characteristics of each array element in the proposed ERA can be flexibly reconfigured into various states, introducing electromagnetic degrees of freedom to enhance wireless system performance. Based on the proposed ERA design, the corresponding channel model is developed. Finally, full-wave simulations are conducted to validate the overall design concept. The results reveal that a gain enhancement of 2.5 dB is achieved at a beamforming direction.

\end{abstract}


\section{Introduction}
Fluid antenna systems (FAS) have recently emerged as a promising technology for enhancing wireless communication, particularly in next-generation (6G) networks with high data rate requirements~\cite{NWK2024FAS_Tutorial}. FAS introduces a new degree of freedom by enabling dynamic control over the activation of antenna array element ports, thereby achieving spatial diversity~\cite{WKK2021First_Paper_on_FAS}. Furthermore, FAS can be potentially integrated with other cutting-edge technologies, such as reconfigurable intelligent surfaces~\cite{Wang_ruiqi2024TAP}, and integrated sensing and communications (ISAC)\cite{LA2022ISAC_Tutorial}, to further enhance future wireless networks.

The majority of investigated FAS works in the literature concentrate on spatial reconfigurability~\cite{WKK2020FAS_limit, Wang_chao2024AI-empowered}. In other words, the position or port of an antenna array can be changed while the intrinsic electromagnetic (EM) properties of each array element are maintained. In such a case, it shares the same concept as movable antennas (MA)~\cite{Zhu_lipeng2024TWC_MA, zhu2024historicalreviewfluidantenna}. However, beyond controlling antenna position, the inherent EM radiation properties can also be reconfigured by reshaping the metallic patterns while maintaining a fixed antenna position\cite{Song_lingnan2019TAP}. This introduces an additional degree of freedom in the EM radiation characteristics of each array element. To more accurately describe this concept, we define it as electromagnetically reconfigurable antennas (ERA), which consist of multiple reconfigurable antenna elements in an array configuration. This can also be referred to as an ``element-reconfigurable array,'' highlighting the unique capability of each array element to undergo electromagnetic reconfigurability. It should be noted that each array element should be around half-wavelength while the majority of existing reconfigurable antenna hardware prototypes are single antennas which cannot constitute antenna arrays due to the very large physical size~\cite{Yujie2022TAP}. Therefore, the wireless communication model based on such hardware is not practically applicable~\cite{liu2025tri-beamforming}. Regarding the ERA, EM reconfigurability can be categorized into three main types: frequency, polarization, and radiation pattern~\cite{Rodrigo2014TAP,Castellanos2025Embracing}.

In this work, we present a practical ERA design for wireless communication systems, where the radiation pattern of each array element can be reconfigured. Based on the ERA design, a single-user MIMO system model is developed, and the corresponding channel model is constructed. To validate the designed ERA, full-wave simulations have been conducted in comparison with a conventional antenna array. The simulation results demonstrate superior beamforming performance of the ERA model at different beamforming directions.

\section{Electromagnetically Reconfigurable Antennas Design}\label{sec:design}

\begin{figure}[t]
  \centering
  \includegraphics[width=0.9\columnwidth]{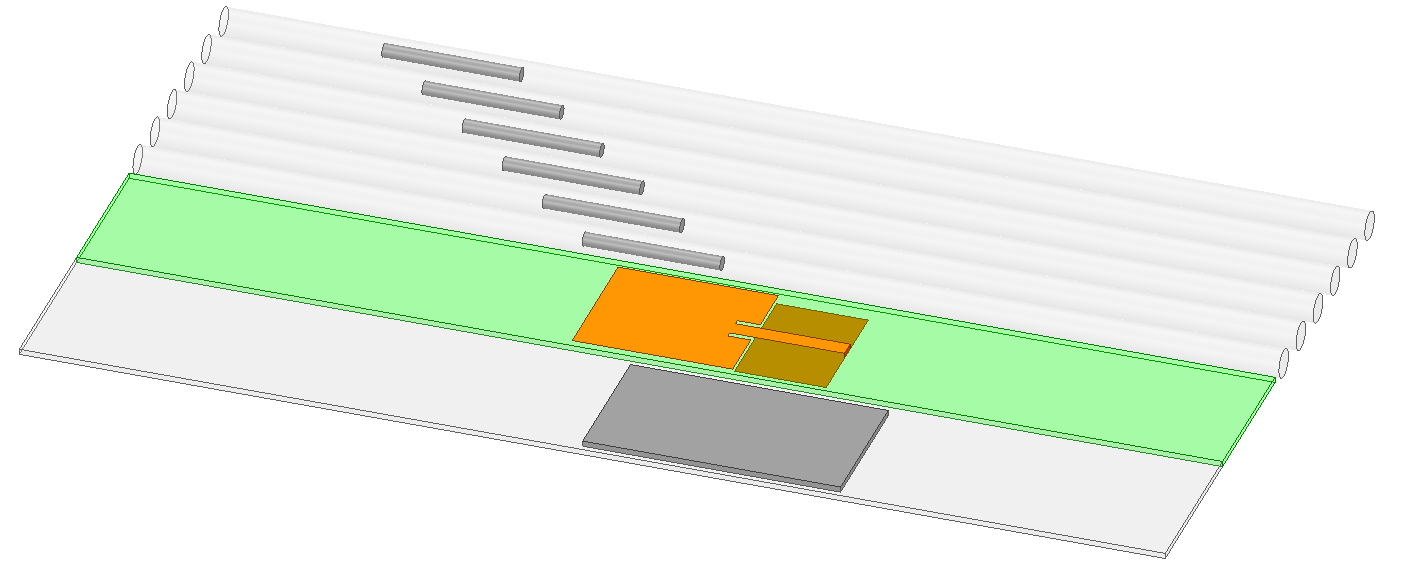}
  \caption{
    The ERA array element design.
  }
  \label{fig_FAS_Array_Element}
\end{figure}

The proposed practical ERA element design is shown in Fig.~\ref{fig_FAS_Array_Element}. The design is inspired by the Yagi-Uda antenna\cite{Yagi1928, HDLu2012Quasi-Yagi}. Specifically, it consists of three main components. The first part includes six parallel cylindrical pipes filled with liquid metal inside the microtubes. Ideally, the liquid metal can be continuously controlled in terms of both length and position using software-controllable microfluidics, where the length and position of each cylindrical fluid element are determined by the amount of metal filled and the air pressure. The entire microtube structure is made of transparent polymethyl methacrylate (PMMA) for ease of visualizing various reconfigurable states. The middle part of the design consists of a planar monopole antenna, which serves as the excitation source for the overall array element. It is composed of double-layered FR4 substrates with metallic patterns for the top monopole and bottom ground. Lastly, another liquid metal element is positioned at the rear of the structure to further enhance the radiation pattern control of the array element. The shape and position of this liquid metal element can be adjusted through a pumping system, considering its larger size. The container is also made of transparent PMMA.

\begin{figure}[t]
  \centering 
  \includegraphics[width=0.7\columnwidth]{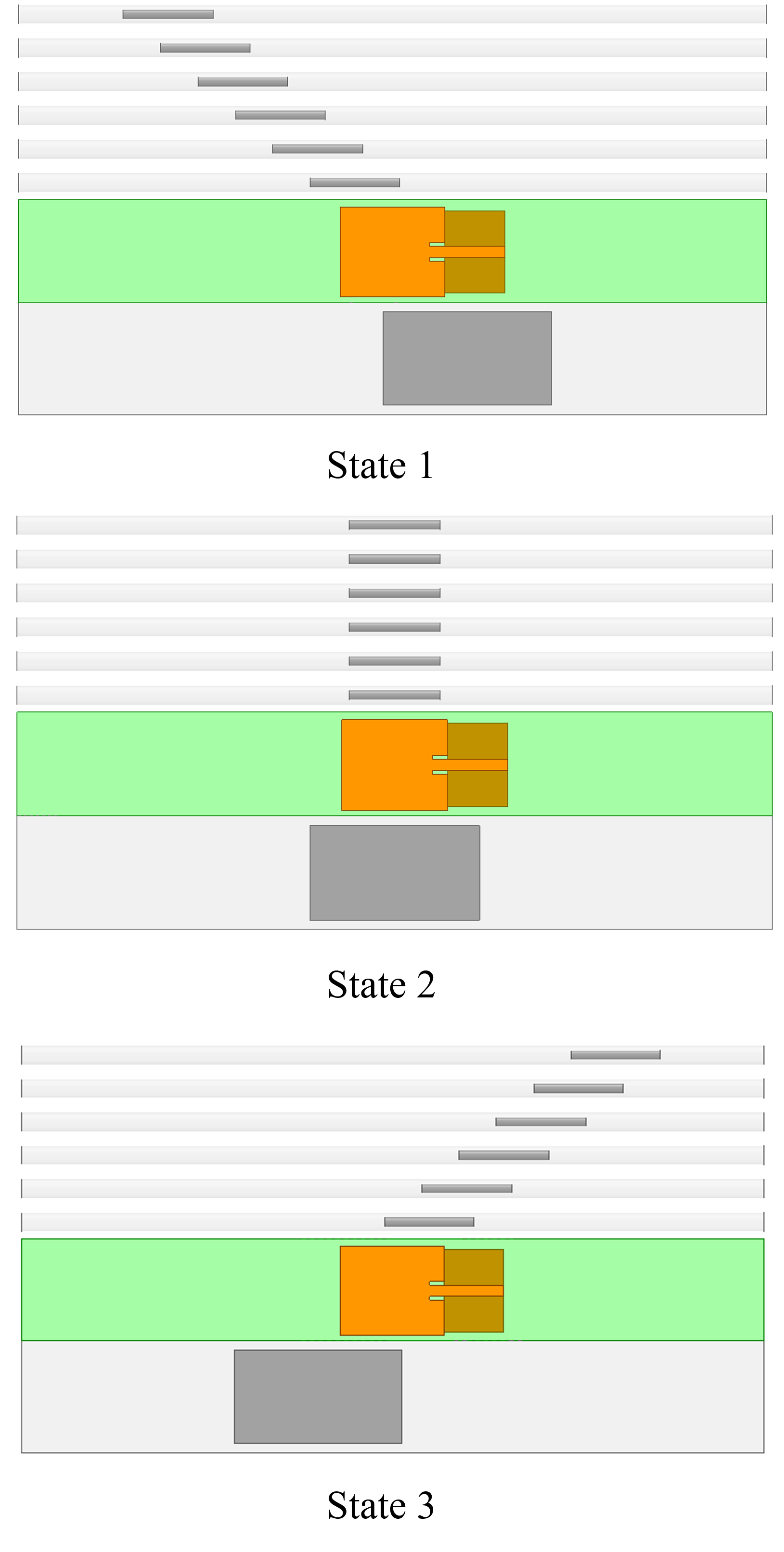}
  \vspace{-1em}
  \caption{
    The operation principle of the ERA array element.
  }
  \label{fig_FAS_Element_Mechanism}
\end{figure}

The operation principle of the reconfigurability is illustrated as follows. The six channels function as directors, while the bulky back parasitic element acts as the reflector in the Yagi-Uda antenna. Here, the number of directors can be reduced to lower the cost and achieve a more compact overall element structure. The number of directors is a trade-off between gain and overall cost. The middle planar monopole antenna serves as the excitation source for the ERA element. By controlling the length and position of both the directors and the reflector, the overall antenna radiation pattern can be altered with continuous beam scanning capability. To better understand the operation principle of the ERA element, three special reconfigurable states are demonstrated in Fig.~\ref{fig_FAS_Element_Mechanism}. These three states feature various director and reflector configurations that reshape the overall radiator structure. State 1 has the director shifted to the left side with the reflector moved to the right side. Consequently, the overall radiation pattern directs the main beam to the left. A similar operation principle applies to States 2 and 3. However, it should be noted that each fluid metal element can take arbitrary shapes and positions, providing extensive freedom in controlling the radiation pattern.

\begin{figure}[t]
  \hspace{\fill} 
  \includegraphics[width=\columnwidth]{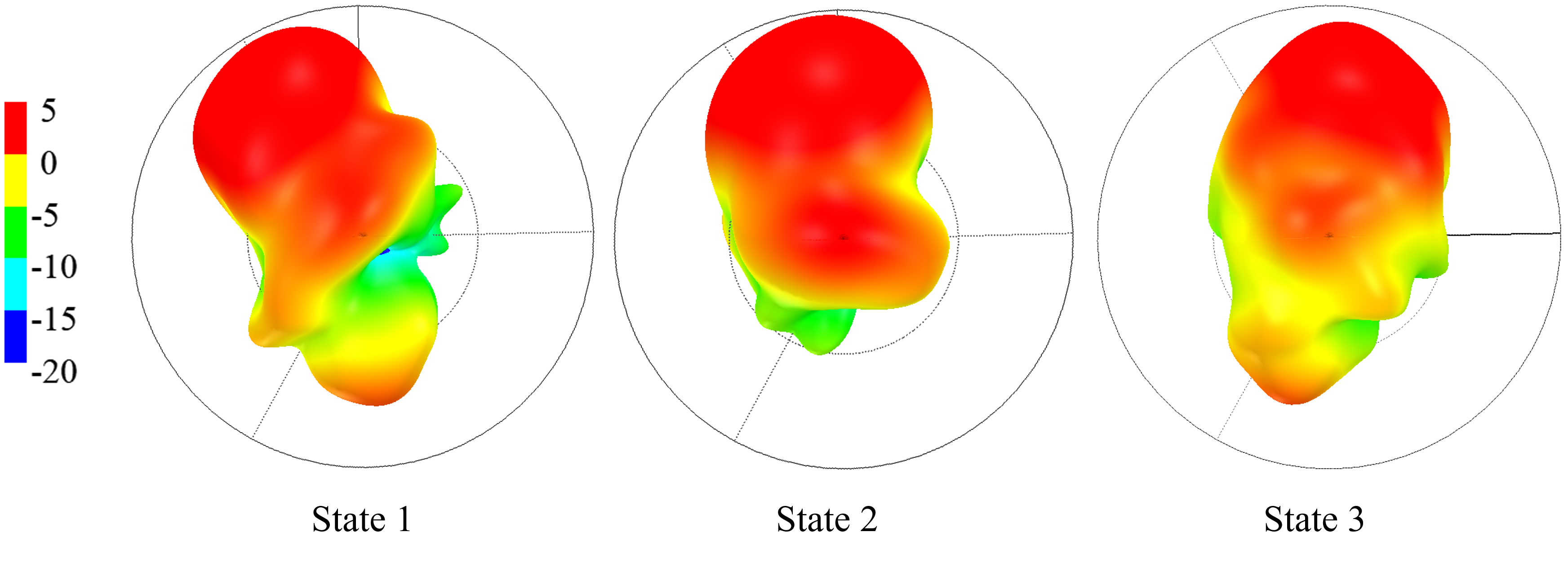}
  \vspace{-1.8em}
  \caption{
    The full-wave simulated radiation patterns of the EAR element at different reconfigurable states.
  }
  \label{fig_FAS_Element_Pattern}
\end{figure}

To validate the overall ERA design, the element is simulated using ANSYS High-Frequency Structure Simulator (HFSS), and the simulated three-dimensional (3D) radiation patterns are shown in Fig.~\ref{fig_FAS_Element_Pattern}. From the simulation results, three distinct main beam radiation patterns are observed, each directed towards different directions. It should be noted that although the distributions of the directors and reflector are symmetric between States 1 and 3, the asymmetric radiation patterns arise due to the relative positions of the directors and reflector to the planar monopole antenna.

\begin{figure*}[t]
  \centering
  \includegraphics[width=\linewidth]{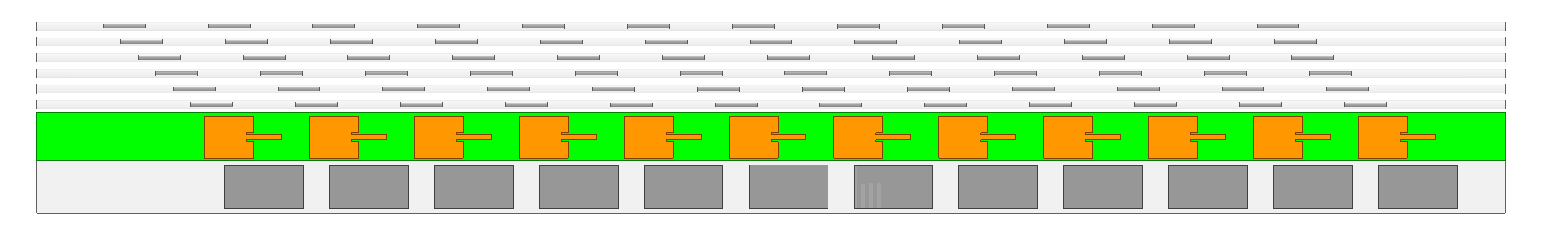}
  \vspace{-1.8em}
  \caption{ 
     Array configuration of the EAR.
    }
  \label{fig_FAS_Array}
\end{figure*}

Based on the designed ERA element, the entire ERA array can be constructed. Due to the structure of the ERA element, a one-dimensional (1D) array can be formed with multiple configurations. In this work, a uniform linear array with 12 elements is constructed, as shown in Fig.~\ref{fig_FAS_Array}. For the ERA array configuration, there are extensive degrees of freedom, including the excitation magnitudes and phases of the planar monopole, as well as the continuous distribution of both the directors and reflectors. These provide substantial possibilities to enhance wireless communication systems. Possible practical applications include, but are not limited to, improved near-field beam focusing and far-field large-angle beam scanning.

\section{ERA-Based EM-Domain Channel Modeling}\label{sec:channel modeling}

To accurately account for the element radiation pattern reconfigurability of the proposed ERAs, this section presents a new wireless communication channel model in the \ac{EM}-domain. The same model has been adopted and extended in~\cite{Journal_Version} for further validation and analysis. 

\subsection{Signal Model}

\begin{figure*}[t]
  \centering
  \includegraphics[width=0.9\linewidth]{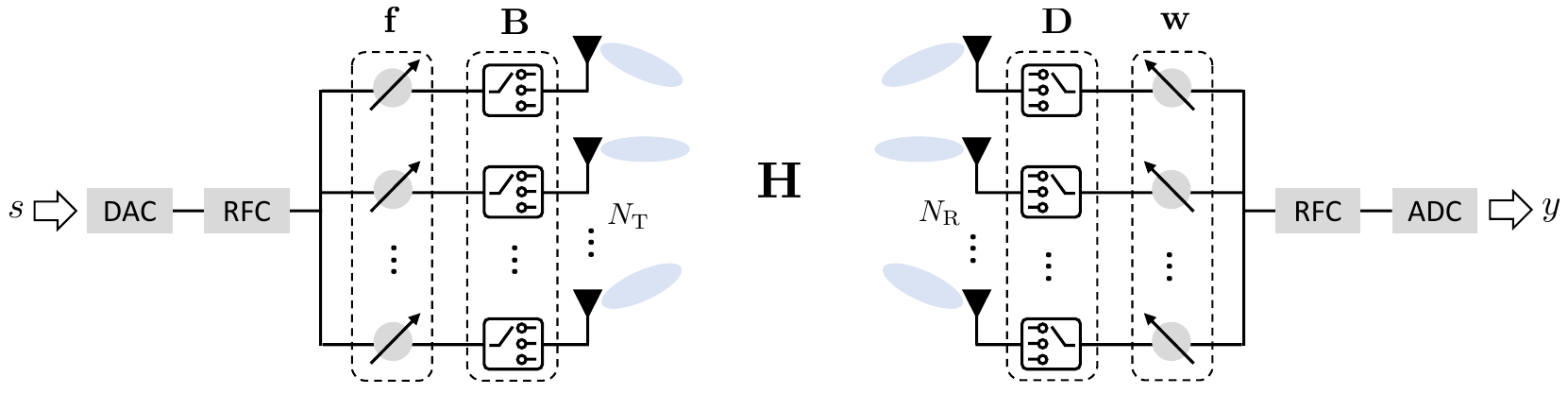}
  \caption{ 
     Simplified hardware block diagram of a single-user \ac{MIMO} system based on ERAs and \ac{RF} phase shifters.
    }
  \label{fig_MIMOsystem}
\end{figure*}

We consider a single-user communication system based on ERAs as demonstrated in Fig.~\ref{fig_MIMOsystem}, where a $N_\Tt$-antenna transmitter (Tx) communicates a single data stream to a $N_\Rt$-antenna receiver (Rx). Both the transmitter and the receiver are equipped with a single \ac{RF} chain. Let~$s\in\mathbb{C}$ denote the transmit symbol. We assume $|s|^2=P_\mathrm{T}$, where $P_\Tt$ represents the transmit power. The transmitter first up-converts the symbol to the carrier frequency by passing through the \ac{RF} chain and then applying an \ac{RF} precoder~$\fv\in\mathbb{C}^{N_\Tt\times 1}$. This \ac{RF} precoder is implemented using analog phase shifters (i.e., the feeding network) with constraint~$|f_i|^2 = 1/N_\Tt$, $i=1,2,\dots,N_\Tt$. After the \ac{RF} precoder, the signal is radiated out via the~$N_\Tt$ transmit antennas, passes through the wireless channel~$\Hm\in\mathbb{C}^{N_\Rt\times N_\Tt}$, and is received by the~$N_\Rt$ receive antennas. Note that in our proposed antenna design, various radiation patterns can be chosen for each antenna individually. This reconfigurability will be reflected in the expression of channel matrix~$\Hm$ later.

The receiver applies a \ac{RF} combiner~$\wv\in\mathbb{C}^{N_\Rt\times 1}$ and then down-converts the radio signal to the baseband through the \ac{RF} chain. Similarly, we constrain the combiner as $|w_j|^2=1/N_\Rt$, $j=1,2,\dots,N_\Rt$. Hence, the final received baseband signal is given by \begin{equation}\label{eq:BBsignal}
    y = \wv^\HH\Hm\fv s + \wv^\HH\nv,
\end{equation}
where~$\nv\sim\mathcal{CN}(\mathbf{0},\sigma^2\mathbf{I}_{N_\Rt})$ denotes additive white Gaussian noise. 

\subsection{Channel Model}
In this work, we target the channel matrix~$\Hm$ in the far-field condition, which is characterized by the \emph{planar wave model}. We will show that our proposed reconfigurable antenna design introduces extra degrees of freedom to reshape the wireless channel. We start by analyzing the wireless channel based on the conventional antenna array and then extend to the proposed ERA. 

\subsubsection{Conventional Antenna Array}
We consider a scenario where a \ac{LoS} and multiple \ac{NLoS} paths exist. Based on the \ac{SV} multipath model, the \ac{NLoS} channel is modeled as the sum of the contributions of $C$ scattering clusters, with the $c^\text{th}$ cluster contributing $L_c$ propagation paths. Assuming an ideal Dirac pulse-shaping filter, the equivalent frequency-domain baseband channel based on a conventional antenna array can be expressed as~\cite{Ayach2014Spatially,Tarboush2021TeraMIMO}
\begin{multline}\label{eq:HFF_conv}
    \Hm_\mathrm{CV} = \gamma\alpha_{\mathrm{LoS}}G_\Rt(\phiv_\mathrm{LoS})G_\Tt(\thetav_\mathrm{LoS})\av_\Rt(\phiv_\mathrm{LoS})\av_\Tt^\HH(\thetav_\mathrm{LoS}) \\
    +\gamma\sum_{c=1}^C\sum_{\ell=1}^{L_c}\alpha_{c,\ell}G_\Rt(\phiv_{c,\ell})G_\Tt(\thetav_{c,\ell})\av_\Rt(\phiv_{c,\ell})\av_\Tt^\HH(\thetav_{c,\ell}),
\end{multline}
where~$\gamma$ is a normalization factor such that~$\gamma = \sqrt{\frac{N_\Tt N_\Rt}{1+\sum_{c=1}^C L_c}}$. Here, $\alpha$ denotes complex channel gain, $\thetav$ denotes \ac{AoD} at the transmitter, and $\phiv$ denotes \ac{AoA} at the receiver, corresponding to different paths. Note that each angle contains an azimuth and elevation components, i.e., $\thetav = [\theta_{\mathrm{az}},\theta_{\mathrm{el}}]^\TT$, $\phiv = [\phi_{\mathrm{az}},\phi_{\mathrm{el}}]^\TT$. For clarity, the azimuth angle is defined as the angle between the positive $X$-axis and the target's projection on the $XOY$-plane, while the elevation angle is the angle between the $Z$-axis and the target direction. Both are expressed in the transmitter's and receiver's body frames. In addition, $G_\Tt$ and $G_\Rt$ represent the transmit and receive \emph{antenna gains}, which are functions of corresponding \ac{AoD} and \ac{AoA}. Finally, the vectors $\av_\Tt$ and $\av_\Rt$ are the normalized receive and transmit \acp{ARV}. Taking $\av_\Tt$ as an example, assuming an $N_\Tt^{\mathrm{h}}\times N_\Tt^{\mathrm{v}}$ \ac{UPA} configuration (thus $N_\Tt=N_\Tt^{\mathrm{h}} N_\Tt^{\mathrm{v}}$), the transmit \ac{ARV} is written as~\cite{Heath2016Overview,Zheng2024Mutual}
\begin{equation}
    \av_\Tt(\thetav) = \frac{1}{\sqrt{N_\Tt}}e^{-j2\pi\theta^{\mathrm{h}}\kv(N_\Tt^\mathrm{h})}\!\otimes\! e^{-j2\pi\theta^\mathrm{v}\kv(N_\Tt^\mathrm{v})},
\end{equation}
where~$\kv(N)=[0,1,\dots,N-1]^\TT$, and~$\theta^\mathrm{h}$ and~$\theta^\mathrm{v}$ are the spatial angles corresponding to the horizontal and vertical dimensions, respectively. Assuming the \ac{UPA} is deployed on the $YOZ$-plane of the transmitter's body coordinate system, we obtain $\theta^\mathrm{h}\triangleq{d_\It}\sin(\theta_\mathrm{az})\sin(\theta_\mathrm{el})/{\lambda}$ and $\theta^\mathrm{v}\triangleq{d_\It}\cos(\theta_\mathrm{el})/{\lambda}$, where~$\lambda$ is the wavelength of the operating frequency and $d_\It$ is the inter-element spacing of the transmit antenna array. The receive \ac{ARV} $\av_\Rt(\phiv)$ is defined in the same way.

\subsubsection{Electromagnetically Reconfigurable Antennas}
In conventional antenna arrays, it is trivial to see that each antenna element shares the same radiation pattern, which is characterized by antenna gain function~$G_\Tt(\thetav)$/$G_\Rt(\phiv)$. 
When using the element-reconfigurable array, these antenna elements can choose different radiation patterns by adjusting the state of the liquid conducting material as demonstrated in Section~\ref{sec:design}. We use~$G_{\Tt,i}(\thetav)$ and $G_{\Rt,j}(\phiv)$ to denote the radiation patterns of the~$i^\text{th}$ transmit antenna and the~$j^\text{th}$ receive antenna, respectively. Each of these radiation patterns is chosen from a preset set of $N$ available radiation patterns, which we denote as~$\{\bar{G}_1,\bar{G}_2,\dots,\bar{G}_N\}$. Defining a dictionary vector $\bar{\gv}(\varphiv) \triangleq [\bar{G}_1(\varphiv),\bar{G}_2(\varphiv),\dots,\bar{G}_N(\varphiv)]^\TT$, we can write
\begin{align}
    G_{\Tt,i}(\thetav) &= \bar{\gv}(\thetav)^\TT \bv_{\Tt,i}, \quad i=1,2,\dots, N_\Tt,\\
    G_{\Rt,j}(\phiv) &= \bar{\gv}(\phiv)^\TT \bv_{\Rt,j}, \quad j=1,2,\dots, N_\Rt.
\end{align}
Here, $\bv_{\Tt,i}$ and $\bv_{\Rt,j}$ are two binary vectors denoting the selection of radiation pattern, which are constrained as 
\begin{equation}\label{eq:ConsSelection}
    \bv_{\Tt,i},\bv_{\Rt,j}\in\{\bv|b_n\in\{0,1\},n=1,2,\dots,N,\|\bv\|_2=1\}.
\end{equation}
For notational convenience, we assume that the transmit and receive antennas share the same set of reconfigurable radiation patterns.

Based on this radiation pattern selection mechanism, we can extend~\eqref{eq:HFF_conv} to our ERA as
\begin{align}\label{eq:HFF_new1}
    &\hspace{-0.5em}\Hm_\mathrm{ER}\! =\! \gamma\alpha_\mathrm{LoS}\big(\gv_\Rt(\phiv_\mathrm{LoS})\!\odot\!\av_\Rt(\phiv_\mathrm{LoS})\big)  \big(\gv_\Tt(\thetav_\mathrm{LoS})\!\odot\!\av_\Tt(\thetav_\mathrm{LoS})\big)^\HH\notag\\
    &\hspace{-0.5em}+ \!\gamma\!\sum_{c=1}^C\!\sum_{\ell=1}^{L_c}\!\!\alpha_{c,\ell}\!\big(\gv_\Rt(\phiv_{c,\ell})\!\odot\!\av_\Rt(\phiv_{c,\ell})\big) \!   \big(\gv_\Tt(\thetav_{c,\ell})\!\odot\!\av_\Tt(\thetav_{c,\ell})\big)^\HH\!\!\!,\!
\end{align}
where $\odot$ denotes the Hadamard product, and $\gv_\Xt(\varphiv)=[G_{\Xt,1}(\varphiv),G_{\Xt,2}(\varphiv),\dots,G_{\Xt,N_{\Xt}}(\varphiv)]^\TT$ for $\Xt\in\{\Tt,\Rt\}$. We further define the following two selection matrices $\Bm$ and $\Dm$ for the transmitter and receiver, respectively:
\begin{align}
    \Bm &= \text{blkdiag}\{\bv_{\Tt,1}^\TT,\bv_{\Tt,2}^\TT,\dots,\bv_{\Tt,N_\Tt}^\TT\}\in\mathbb{R}^{N_\Tt\times NN_\Tt},\label{eq:Bdef}\\
    \Dm &= \text{blkdiag}\{\bv_{\Rt,1}^\TT,\bv_{\Rt,2}^\TT,\dots,\bv_{\Rt,N_\Rt}^\TT\}\in\mathbb{R}^{N_\Rt\times NN_\Rt}.\label{eq:Ddef}
\end{align}
Then, we have
\begin{align}  \gv_\Tt(\thetav)\odot\av_\Tt(\thetav)&=\Bm\big(\av_\Tt(\thetav)\otimes\bar{\gv}(\thetav)\big),\label{eq:gtat}\\
\gv_\Rt(\phiv)\odot\av_\Rt(\phiv)&=\Dm\big(\av_\Rt(\phiv)\otimes\bar{\gv}(\phiv)\big),\label{eq:grar}
\end{align}
where $\otimes$ denotes the Kronecker product. 
Substituting \eqref{eq:gtat} and~\eqref{eq:grar} into \eqref{eq:HFF_new1}, we obtain
\begin{equation}\label{eq:HERFF_2}
    \Hm_\mathrm{ER} = \gamma\Dm{\Hm}_\mathrm{EM}\Bm^\TT,
\end{equation}
where~${\Hm}_\mathrm{EM}\in\mathbb{C}^{NN_\Rt\times NN_\Tt}$ is called the \ac{EM}-domain channel~\cite{Ying2024Reconfigurable} given by
\begin{multline}\label{eq:tildeHER}
    {\Hm}_\mathrm{EM} = \alpha_\mathrm{LoS}\big(\av_\Rt(\phiv_\mathrm{LoS})\!\otimes\!\bar{\gv}(\phiv_\mathrm{LoS})\big)\big(\av_\Tt(\thetav_\mathrm{LoS})\!\otimes\!\bar{\gv}(\thetav_\mathrm{LoS})\big)^\HH\\
    \!+ \!\sum_{c=1}^C\!\sum_{\ell=1}^{L_c}\!\alpha_{c,\ell}\big(\av_\Rt(\phiv_{c,\ell})\!\otimes\!\bar{\gv}(\phiv_{c,\ell})\big)  \big(\av_\Tt(\thetav_{c,\ell})\!\otimes\!\bar{\gv}(\thetav_{c,\ell})\big)^\HH\!\!\!.
\end{multline}
The matrices $\Bm$ and $\Dm$ fully describe the radiation pattern configurations of all antennas at the transmitter and receiver, as depicted in Fig.~\ref{fig_MIMOsystem}.

\subsection{Beampattern Synthesis}

Based on the derived channel model, one can compute the far-field array beampattern (power intensity) as~\cite{Zheng_pinjun2024MC,Zheng2025Enhanced}
\begin{equation}\label{eq:Etheta}
    E(\thetav) = \big|\big(\av_\mathrm{T}(\thetav)\otimes\bar{\gv}(\thetav)\big)^\HH\Bm^\TT\fv\big|^2.
\end{equation}
Using this beampattern synthesis calculation equation, we can evaluate the accuracy of the derived channel model by comparing the beampatterns computed via~\eqref{eq:Etheta} and that obtained through full-wave HFSS simulation. We will show these results in Section~\ref{sec:Full-wave simulation}.

\section{Simulation Results}\label{sec:Full-wave simulation}

\begin{figure*}[t]
  \centering
  \includegraphics[width=0.9\linewidth]{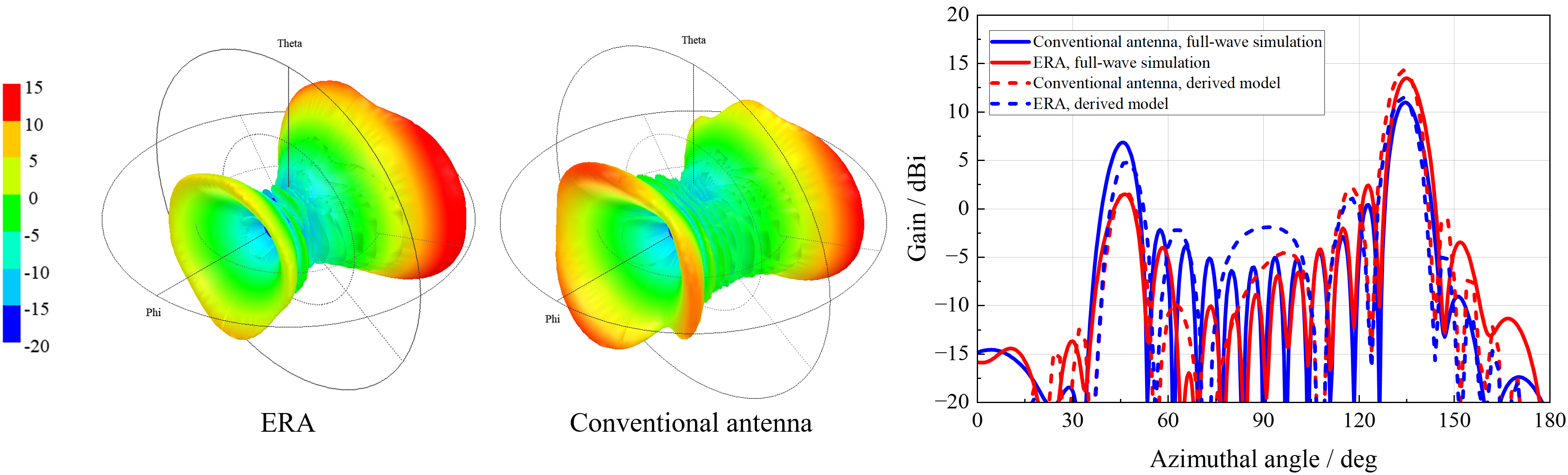}
  \caption{ 
    The simulated 3D radiation patterns for beamforming to 135${^\circ}$ of the designed ERA compared with the conventional antenna. 
   }
\label{fig_RPs_m90}
\end{figure*}

The full-wave electromagnetic (EM) simulation of the designed ERA array configuration (shown in Fig.\ref{fig_FAS_Array}) is conducted using ANSYS HFSS. The beamforming direction with angles of 135${^\circ}$ is simulated, where the phase differences between adjacent array elements is 180${^\circ}$. For comparison with a conventional antenna array, an array with a fixed element radiation pattern is selected as the benchmark, where the element radiation pattern is defined as state2 in Fig.~\ref{fig_FAS_Element_Pattern}.

The simulated radiation patterns of the conventional antenna and the designed ERA is shown in Fig.\ref{fig_RPs_m90}. From the full-wave simulated 3D radiation patterns, it can be observed that the overall radiation patterns exhibit a higher realized gain at the desired main beam angle compared to the conventional antenna. It should be noted that the feeding phases for the ERA and conventional antennas are the same for a fair comparison.

To quantitatively illustrate the superior performance of the designed ERA, 2D radiation patterns are also included in Fig.\ref{fig_RPs_m90}. For the beamforming angle of 135${^\circ}$, the designed ERA demonstrates an overall gain of 13.5 dBi, which is 2.5 dB higher than the conventional antenna. Meanwhile, the main side lobe at 45${^\circ}$ is suppressed from 6.7 dBi for the conventional antenna to 1.2 dBi for the ERA. Therefore, a side lobe reduction of 5.5 dB is achieved. Moreover, good agreement is observed between calculated results from the derived model and full-wave simulation results. Thus, the ERA design and concept have been validated through full-wave EM simulation.

\section{Conclusion}\label{sec:Conclusion}

In this work, an electromagnetically reconfigurable antenna (ERA) design is presented for future advanced wireless communication systems, where each ERA array element can be electromagnetically reconfigured. This introduces an additional degree of electromagnetic freedom in the MIMO system, and a corresponding channel model is established. The entire ERA design has been validated through full-wave EM simulation. Finally, this work focuses solely on the radiation pattern reconfigurability of each array element, while the frequency and polarization features can be explored in future studies.

\bibliography{references}
\bibliographystyle{IEEEtran}

\end{document}

%% file: preamble.tex
\usepackage{amsmath,amsfonts}
\usepackage{array}
\usepackage[caption=false,font=normalsize,labelfont=sf,textfont=sf]{subfig}
\usepackage{textcomp}
\usepackage{stfloats}
\usepackage{verbatim}
\usepackage{graphicx}
\usepackage{cite}
\usepackage{tabularx}
\usepackage{booktabs}

\usepackage{acronym}
\acrodef{MIMO}{multiple-input multiple-output}
\acrodef{RF}{radio frequency}
\acrodef{LoS}{line-of-sight}
\acrodef{NLoS}{non-line-of-sight}
\acrodef{AoA}{angle-of-arrival}
\acrodef{AoD}{angle-of-departure}
\acrodef{UPA}{uniform planar array}
\acrodef{ARV}{array response vector}
\acrodef{EM}{electromagnetic}
\acrodef{ER-FAS}{electromagnetically reconfigurable fluid antenna system}
\acrodef{SV}{Saleh-Valenzuela}

\usepackage{color}

\usepackage{tikz} 
\usepackage[utf8]{inputenc}
\usepackage{pgfplots}
\usepackage{tabu,longtable}
\usepackage{diagbox}
\usepackage{threeparttable}
\usepackage{makecell}
\usepackage{multirow} 
\usepackage[bottom=1.04in,top=0.75in,left=0.625in,right=0.625in]{geometry}
\usepackage{units} 		
\usepackage{bm} 		
\usepackage{amssymb} 	
\include{symbol}        
\usepackage{hyperref} 	
\usepackage{amsmath}

\usepackage{algorithm}
\usepackage{algpseudocode}
\algtext*{EndWhile}
\algtext*{EndIf}
\algtext*{EndFor}

\makeatletter
\algnewcommand{\LineComment}[1]{\Statex \hskip\ALG@thistlm \(\triangleright\) #1}
\makeatother

%% file: symbol.tex
\newcommand{\TT}{\mathsf{T}}
\newcommand{\HH}{\mathsf{H}}

\newcommand{\av}{{\bf a}}
\newcommand{\bv}{{\bf b}}

\newcommand{\fv}{{\bf f}}
\newcommand{\gv}{{\bf g}}

\newcommand{\kv}{{\bf k}}

\newcommand{\nv}{{\bf n}}

\newcommand{\wv}{{\bf w}}

\newcommand{\Bm}{{\bf B}}

\newcommand{\Dm}{{\bf D}}

\newcommand{\Hm}{{\bf H}}

\newcommand{\It}{{\rm I}}

\newcommand{\Rt}{{\rm R}}

\newcommand{\Tt}{{\rm T}}

\newcommand{\Xt}{{\rm X}}

\newcommand{\phiv}{\hbox{\boldmath$\phi$}}
\newcommand{\varphiv}{\hbox{\boldmath$\varphi$}}

\newcommand{\thetav}{\hbox{$\boldsymbol\theta$}}